\documentclass[useAMS,usenatbib,usegraphicx]{mn2e}

\newcommand {\vthp}		{v_\mathrm{th,p}}
\newcommand {\Vsh}		{V_\mathrm{sh}}
\newcommand {\nel}		{n_\mathrm{e}}
\newcommand {\np}		{n_\mathrm{p}}
\newcommand {\nH}		{n_\mathrm{H}}
\newcommand {\mpr}		{m_\mathrm{p}}
\newcommand {\me}		{m_\mathrm{e}}
\newcommand {\Bsat}		{B_\mathrm{sat}}

\newcommand {\refp}[1]	{(\ref{#1})}
\newcommand {\ns}		{n_\mathrm{s}}
\newcommand {\qs}		{q_\mathrm{s}}
\newcommand {\ms}		{m_\mathrm{s}}
\newcommand {\sigmas}	{\sigma_\mathrm{s}}
\newcommand {\alphas}	{\alpha_\mathrm{s}}
\newcommand {\omegaps}	{\omega_\mathrm{ps}}
\newcommand {\zetas}	{\zeta_\mathrm{s}}
\newcommand {\kmax}		{k_\mathrm{max}}

\title[A Possible Origin of Magnetic Fields] 
{A Possible Origin of
Magnetic Fields in Galaxies and Clusters: Strong Magnetic fields at
$z\sim 10$?}  \author[Y. Fujita and T. N. Kato] {Yutaka
Fujita$^{1,2}$\thanks{E-mail: yfujita@th.nao.ac.jp (YF);
tkato@th.nao.ac.jp (TNK)} and Tsunehiko N.  Kato$^{1}$\footnotemark[1]
\\
$^{1}$National Astronomical Observatory, Osawa 2-21-1,
Mitaka, Tokyo 181-8588, Japan\\
$^{2}$Department
of Astronomical Science, The Graduate University for Advanced Studies,
Osawa 2-21-1, Mitaka, Tokyo 181-8588, Japan}
\begin{document}

\date{Accepted 0000 December 00. Received 0000 December 00; in original
form 0000 October 00}

\pagerange{\pageref{firstpage}--\pageref{lastpage}} \pubyear{0000}

\maketitle

\label{firstpage}

\begin{abstract}
We propose that strong magnetic fields should be generated at shock
waves associated with formation of galaxies or clusters of galaxies by
the Weibel instability, an instability in collisionless plasmas. The
strength of the magnetic fields generated through this mechanism is
close to the order of those observed in galaxies or clusters of galaxies
at present. If the generated fields do not decay rapidly, this indicates
that strong amplification of magnetic fields after formation of galaxies
or clusters of galaxies is not required.  This mechanism could have
worked even at a redshift of $\sim 10$, and therefore the generated
magnetic fields may have affected the formation of stars in
protogalaxies. This model will partially be confirmed by future
observations of nearby clusters of galaxies.  Mechanisms that preserve
the magnetic fields for a long time without considerable decay are
discussed.
\end{abstract}

\begin{keywords}
instabilities --- magnetic fields --- galaxies: general ---
galaxies: clusters: general
\end{keywords}

\section{Introduction}

The question of the origin of galactic magnetic fields is one of the
most challenging problems in modern astrophysics. It is generally
assumed that magnetic fields in spiral galaxies are amplified and
maintained by a dynamo through rotation of the galaxies \citep{wid02}.
The dynamo requires seed fields to be amplified. However, observations
of microgauss fields in galaxies at moderate redshifts strongly
constrain the lower boundary of the seed fields \citep{ath98}. Moreover,
magnetic fields are also observed in elliptical galaxies and galaxy
clusters, in which rotation cannot play a central role as the dynamo
mechanism \citep*{cla01,wid02,val04}.

The Weibel instability is another mechanism to generate strong magnetic
fields \citep{wei59,fri59}.  This instability is driven in a
collisionless plasma, or a tenuous ionised gas, by the anisotropy of the
particle velocity distribution function (PDF) of the plasma. When the
PDF is anisotropic, currents and then magnetic fields are generated in
the plasma so that the plasma particles are deflected and the PDF
becomes isotropic \citep{med99}. Through the instability, the free
energy attributed to the PDF anisotropy is transferred to magnetic field
energy. This instability does not need seed magnetic fields. It can be
saturated only by nonlinear effects, and thus the magnetic fields can be
amplified to very high values. This instability has been observed
directly in recent laser experiments \citep{wei02}. In astrophysical
plasmas, the instability is expected to develop at shocks or at steep
temperature gradients, where the PDF is anisotropic. Examples of the
sites are pulsar winds, shocks produced by gamma-ray bursts, jets from
active galactic nuclei (AGNs), cosmological shocks, and cold fronts
(contact discontinuities between cold and hot gas) in clusters of
galaxies \citep{med99,kaz98,nis03,sch03,oka03}. Although the instability
was found in 1959, its nonlinear nature had prevented us from
understanding its long-term evolution. Recently, however, as computer
power increases, detailed particle simulations of plasmas have been
initiated and they have revealed the evolution of magnetic fields even
after saturation of the instability \citep{sil03,med05}. Based on these
results, we consider the generation of magnetic fields at galaxy and
cluster-scale shocks through the Weibel instability at the formation of
galaxies (both ellipticals and spirals) and clusters. We use the
cosmological parameters of $\Omega_0=0.3$, $\lambda=0.7$, the Hubble
constant of $H_0=70\rm\: km\: s^{-1}\: Mpc^{-1}$, and $\sigma_8=0.9$.

\section{Models}

\subsection{Generation of Magnetic Fields at Shocks}

At the vicinity of the shock front of a collisionless shock, particles
from the upstream are mixed up with those in the downstream, and an
anisotropy of PDF will be generated in the plasma.  As discussed by
\citet{sch03}, the particles from the upstream region firstly will be
affected by the Langmuir instability. Since the Langmuir instability is
a longitudinal electrostatic mode, the velocity component parallel to
the shock normal will be thermalized, while the other components will
remain unaffected.  Therefore, it is natural to assume that the thermal
velocity parallel to the shock normal is on the order of the relative
velocity between the upstream and the downstream or on the order of the
shock velocity $\Vsh$, and those perpendicular to the shock normal are
on the order of the thermal velocity of the upstream plasma.  This
velocity or temperature anisotropy should develop the Weibel
instability.

For shocks in electron-proton plasmas,
\citet{sch03} indicated that the temperature anisotropy for electrons is
too small to drive the Weibel instability if $\mathcal{M}
\la (\mpr/\me)^{1/2}=43$, where $\mathcal{M} = \Vsh/\vthp$ is the shock
Mach number, $\vthp$ is the proton thermal velocity in the upstream, and
$\mpr$ and $\me$ are the mass of a proton and that of an electron,
respectively.  However, if $\mathcal{M} \ga2$
 (note that $\mathcal{M}$ provides a measure of the anisotropy in
protons), the temperature anisotropy for the protons is large enough to
drive the Weibel instability; this is the case, even if the PDF of the
electrons is completely isotropic (see Appendix \ref{sec:AppendixA}).

The magnetic field strength reaches its maximum when the Weibel
instability saturates, and the saturation level would be given as
follows.  As the Weibel instability develops, magnetic fields are
generated around numerous current filaments \citep{med05,kat05}. The
instability saturates when the generated magnetic fields eventually
interrupt the current in each filament, or in other words when the
particle's gyroradii in the excited magnetic fields are comparable to
the characteristic wavelength of the excited field \citep{med99}.  For
electron-positron plasmas, \citet{kat05} showed that the typical current
at and after the saturation is given by the Alfv\'{e}n current
\citep{alf39}.  In terms of the Alfv\'{e}n current, the saturated
magnetic field strength is given by
\begin{equation}
	\label{eq:Bsat}
	\Bsat \sim \sqrt{2} I_A / (\tilde{R}c),
\end{equation}
where $\tilde{R}$ is the typical radius of a current filament at the
saturation and $I_A \equiv \langle v_{\parallel,\rm e}/c \rangle \me
c^3/e$ is the Alfv\'{e}n current (for nonrelativistic cases).  Here,
$v_{\parallel, \rm e}$ is the electron thermal velocity in the direction
of the higher temperature.  For strongly anisotropic cases \citep[i.e.,
the particle limit cases discussed by][]{kat05}, the radius at the
saturation is given by $\tilde{R}\sim2c/\omega_{\rm pe}$, where
$\omega_{\rm pe}\equiv(4\pi \nel e^2 / \me)^{1/2}$ is the electron
plasma frequency, and the saturated magnetic field strength is given by
\begin{equation}
\label{eq:mag_e} 
	\Bsat \sim 0.5\ v_{\parallel, \rm e}\ (4\pi\nel\me)^{1/2},
\end{equation}
where $\nel$ is the electron number density. 

For electron-proton plasmas we consider below, currents carried by
heavier protons will generate stronger magnetic fields at the
saturation. It is natural to assume that the saturation for the proton
currents is determined by the Alfv\'{e}n current defined for protons,
$I'_A \equiv \langle v_{\parallel, \rm p}/c \rangle \mpr c^3/e$, where
$v_{\parallel, \rm p}$ is the proton thermal velocity in the direction
of the higher temperature.  In a shock wave, this leads to
\begin{equation}
\label{eq:mag} 
\Bsat \sim 0.5\ \Vsh\ (2\pi \np \mpr)^{1/2}\:,
\end{equation}
where $\np$ is the proton number density and we used the assumption of
$v_{\parallel, \rm p} \sim \Vsh$.  This expression is valid when the PDF
of the protons has a strong anisotropy or $\mathcal{M} \ga 2$
\citep{kat05}.  Note that the energy density of the saturated magnetic
fields attains to a sub-equipartition level with the particle kinetic
energy.  The numbers in the parentheses in equations~(\ref{eq:mag_e})
and (\ref{eq:mag}) are different by a factor of two, because in
electron-positron plasmas, both electron and positron compose currents
at the same time, while in electron-proton plasmas, only protons
contribute to currents when the proton Weibel instability develops. In
electron-proton plasmas, if electron anisotropy is sufficiently large,
electrons first generate magnetic fields, which saturate at the early
stage of the instability, and then protons generate magnetic fields,
which leads to the second saturation.  Such an evolution was also
observed in numerical simulations \citep{fre04}.

The magnetic fields generated at the shock front will be convected
downstream at the fluid velocity (in the shock rest frame).  At least
for a short period after the saturation, the magnetic field strength
decreases as the current filaments merge together, because the current
in each filament is limited to the proton Alfv\'{e}n current while the
size of the current filaments increases \citep{kat05}.  However, the
long-term evolution is still an open question.  \citet{med05} indicated
that the field should reach an asymptotic value beyond which it will not
decay, although their model can be applied only when currents are not
dissipated. Recent particle simulations have also shown that the final
magnetic field strength is given by $B_f=\eta_{\rm mer}^{1/2}\Bsat$,
where $\eta_{\rm mer}\sim 0.01$ \citep{sil03,med05}.  In the following,
we assume that the magnetic fields of $B_f$ are preserved.  The very
long-term evolution of the magnetic fields, for which the results of
\citet{med05} and \citet{sil03} cannot be directly applied, will be
discussed in Section~\ref{sec:uns}.

We note that unfortunately, at present, most simulations treat
electron-positron shocks and there are virtually no simulations that
reveal the evolution of collisionless electron-proton shocks
satisfactorily (in terms of box sizes, duration times, and so on),
because of the lack of computational power. Therefore, the model stated
above and used in this paper should be confirmed by direct numerical
simulations in the future.

\subsection{Shock Formation}
\label{sec:shockform}

According to the standard hierarchical clustering scenario of the
universe, an initial density fluctuation of dark matter in the universe
gravitationally grows and collapses; its evolution can be approximated
by that of a spherical uniform over-dense region
\citep{gun72,pee80}. The collapsed objects are called `dark halos' and
the gas in these objects later forms galaxies or clusters of
galaxies. At the collapse, the gas is heated by the `virial shocks' to
the virial temperature of the dark halo, $T_{\rm vir}=G M/(2 r_{\rm
vir})$, where $G$ is the gravitational constant, and $M$ and $r_{\rm
vir}$ are the mass and the virial radius of the dark halo,
respectively. The relation between the virial radius and the virial mass
of an object is given by
\begin{equation}
 \label{eq:r_vir}
r_{\rm vir}=\left[\frac{3\: M}
{4\pi \Delta_c(z) \rho_{\rm crit}(z)}\right]^{1/3}\:,
\end{equation}
where $\rho_{\rm crit}(z)$ is the critical density of the universe, and
$\Delta_c(z)$ is the ratio of the average density of the object to the
critical density at redshift $z$. The critical density depends on
redshift because the Hubble constant depends on that, and it is
given by
\begin{equation}
\label{eq:rho_crit}
 \rho_{\rm crit}(z)
=\frac{\rho_{\rm crit,0}\Omega_0 (1+z)^3}{\Omega(z)}\:,
\end{equation} 
where $\rho_{\rm crit,0}$ is the critical density at $z=0$, and
$\Omega(z)$ is the cosmological density parameter given by
\begin{equation}
 \Omega(z) = \frac{\Omega_0 (1+z)^3}{\Omega_0 (1+z)^3 + \lambda}
\end{equation}
for the flat universe with non-zero cosmological constant. The ratio
$\Delta_c(z)$ is given by
\begin{equation}
\label{eq:Dc_lam}
  \Delta_c(z)=18\:\pi^2+82 x-39 x^2\:, 
\end{equation}
for the flat universe \citep{bry98}, where the parameter $x$ is given by
$x=\Omega(z)-1$. The virial shocks form at $r\approx r_{\rm vir}$ and
the velocity is $v_{\rm vir}\approx \sqrt{G M/r_{\rm vir}}$.

In addition, recent cosmological numerical simulations have shown that
`large-scale structure (LSS) shocks' form even before the collapse
\citep{cen99,min00,dav01,ryu03}. They form at the turnaround radius
($r_{\rm ta}\sim 2 r_{\rm vir}$), the point at which the density
fluctuation breaks off from the cosmological expansion.  For simplicity,
we assume that $r_{\rm ta}= 2 r_{\rm vir}$, which is close to the
self-similar infall solution for a particular mass shell in the
Einstein-de Sitter Universe \citep[$r_{\rm ta}= 1/0.56\: r_{\rm
vir}$;][]{ber85}. The gas that later forms a galaxy or a cluster passes
two types of shocks; first, the gas passes the outer LSS shock, and
then, the inner virial shock. The typical velocity of the LSS shocks is
$\Vsh\approx H(z)r_p$ \citep{fur04}, where $H(z)$ is the Hubble constant
at redshift $z$, and $r_p$ is the physical radius that the region would
have had if it had expanded uniformly with the cosmological
expansion. The temperature of the postshock gas is $T_s\approx 3/16\:
(\mu \mpr/k_\mathrm{B}) \Vsh^2$, where $\mu \mpr$ is the mean particle
mass, and $k_\mathrm{B}$ is the Boltzmann constant. Note that although
the model of \citet{fur04} has been compared with numerical simulations
at $z\sim 0$, it has not been at high-redshifts; it might have some
ambiguity there. We do not consider mergers of objects that have already
collapsed as the sites of magnetic field generation because the Weibel
instability applies only to initially unmagnetised or weakly magnetised
plasmas; at the merger, collapsed objects just bring their magnetic
fields to the newly born merged object.

Since the Weibel instability develops in ionised gas (plasma), we need
to consider the ionisation history of the universe. After the entire
universe is ionised by stars and/or AGNs ($z\la 8$), magnetic
fields are first generated at the LSS shocks. In this case, we do not
consider the subsequent generation of magnetic fields at the inner
virial shocks, because the strength is at most comparable to that of the
magnetic fields generated at the LSS shocks.  On the other hand, when
the universe is not ionised ($z\ga 8$), the Weibel instability
cannot develop at the outer LSS shocks.  However, if the LSS shocks heat
the gas (mostly hydrogen) to $T_s>10^4{\rm\: K}$ and ionise it, the
instability can develop at the inner virial shocks.

In this case, the gas ionised at the LSS shock may recombines before it
reaches the virial shock. The recombination time-scale is given by
\begin{equation}
 \tau_{\rm rec} =\frac{1}{\alpha \nel}
\approx 1.22\times 10^5\; {\rm yr}\:\frac{1}{y}
\left(\frac{T}{\rm 10^4 K}\right)^{0.7}
\left(\frac{\nH}{\rm cm^{-3}}\right)^{-1}
\:,
\end{equation}
where $\alpha$ is the recombination coefficient, $T$ is the gas
temperature, $y$ is the ionisation fraction, and $\nH$ is the hydrogen
density \citep{sha87}. If we assume that $\tau_{\rm rec}=\tau_{\rm
dyn}$, where $\tau_{\rm dyn}\approx (1/2)r_{\rm ta}/\Vsh$ is the
time-scale that the gas moves from the LSS shock to the virial shock,
the ionisation rate when the gas reaches the virial shock is
\begin{equation}
\label{eq:rec}
 y\approx \left(\frac{\tau_{\rm dyn}}
{1.22\times 10^5\; {\rm yr}}\right)^{-1}
\left(\frac{T}{\rm 10^4 K}\right)^{0.7}
\left(\frac{\nH}{\rm cm^{-3}}\right)^{-1}
\;
\end{equation}
for $y<1$. We found that $y<1$ for $z\ga 9$, and the minimum value when
the generation of magnetic fields is effective ($z\la 12$, see
\S\ref{sec:results}) is $y\sim 0.3$. When $y<1$, we simply replace $\np$
in equation~(\ref{eq:mag}) with $y \nH$. For temperature, we assumed
that $T=T_s$ in equation~(\ref{eq:rec}). Since the magnetic fields do
not much depend on the temperature ($\Bsat\propto y^{0.5}\propto
T^{0.35}$), they do not much change even when radiative cooling reduces
the temperature; at least $\Bsat$ does not change by many orders of
magnitude.  It was shown that the ionisation rate just behind a shock is
$y\sim 0.1$, if the shock velocity is relatively small \citep[$\Vsh\sim
40\rm\; km\; s^{-1}$;][]{sha87,sus98}. If the shock velocity is larger,
$y\sim 1$. In our calculations, the velocity of the LSS shocks is
$\Vsh\ga 40\rm\; km\; s^{-1}$, when the generation of magnetic fields is
effective. Thus, $y$ just behind the shocks is at least comparable to
that obtained through the condition of $\tau_{\rm rec}\sim \tau_{\rm
dyn}$, and the incomplete ionisation does not much affect the results
shown in the next section (see also Section~\ref{sec:uns}).

\section{Results and Discussion}
\label{sec:results}

\begin{figure}
\includegraphics[width=84mm]{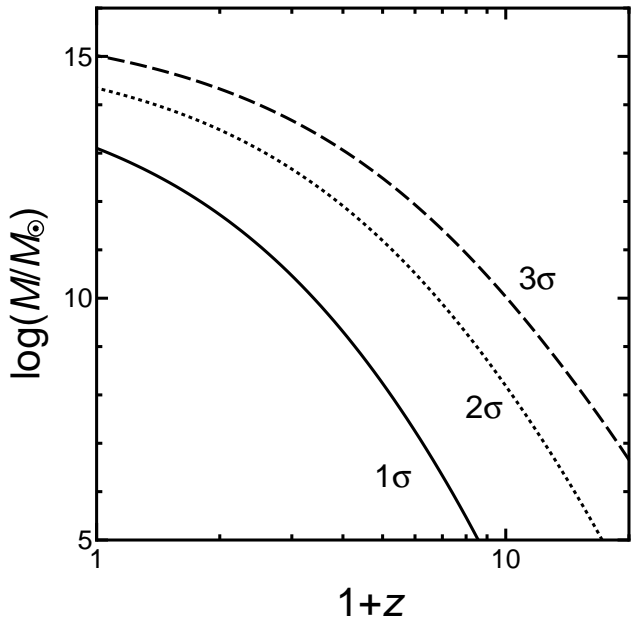}
\caption{Typical masses of objects
forming at redshift $z$. The labels, $1\sigma$, $2\sigma$, and $3\sigma$
(solid, dotted, and dashed lines, respectively) indicate the amplitudes
of initial density fluctuations from which the objects formed; $\sim
1$--$3\sigma$ is the typical value \citep{bar01}. Objects with masses of
$M\la 10^{12}\:M_\odot$ and $M\ga 10^{13}\:M_\odot$ correspond
to galaxies (ellipticals and spirals) and clusters of galaxies,
respectively. }
\label{fig:mass}
\end{figure}

\begin{figure}
\includegraphics[width=84mm]{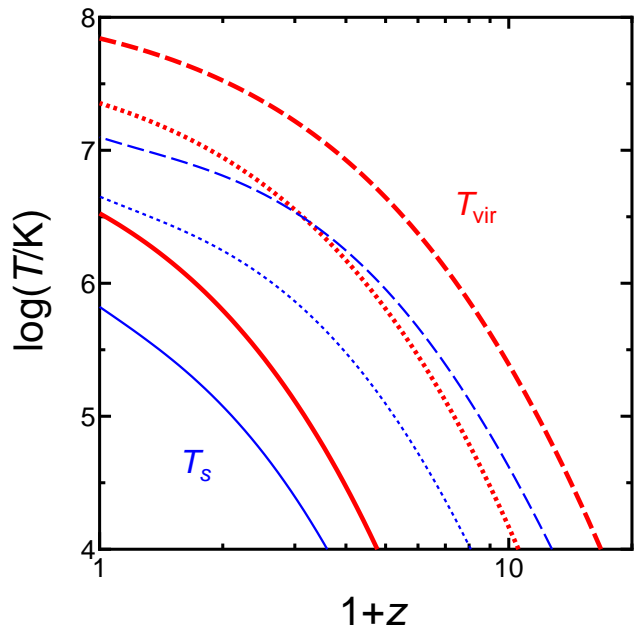}
\caption{Temperatures behind the virial shocks ($T_{\rm
vir}$; thick lines) and those behind the LSS shocks ($T_s$; thin lines)
for objects forming at redshift $z$. Solid, dotted, and dashed lines
correspond to $1\sigma$, $2\sigma$, and $3\sigma$ fluctuations,
respectively (see Fig.\ref{fig:mass})}
\label{fig:temp}
\end{figure}

\begin{figure}
\includegraphics[width=84mm]{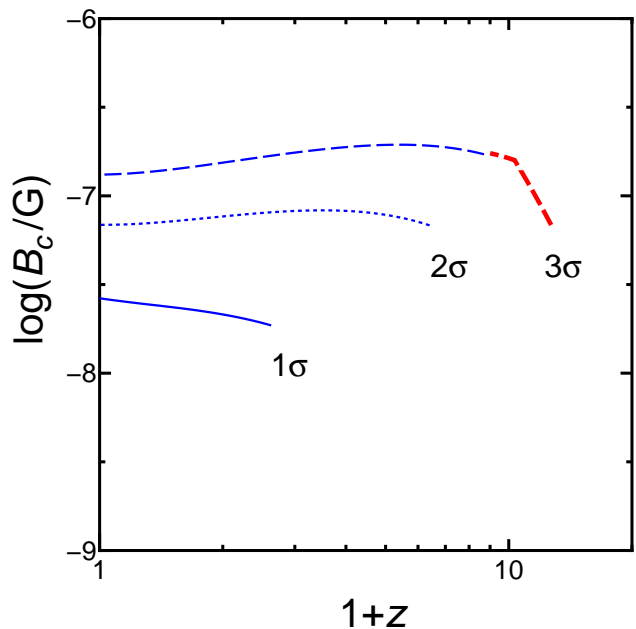} \caption{Typical magnetic field
strengths of objects forming at redshift $z$. Solid, dotted, and dashed
lines correspond to $1\sigma$, $2\sigma$, and $3\sigma$ fluctuations,
respectively (see Fig.\ref{fig:mass}).}  \label{fig:B}
\end{figure}

Fig.\ref{fig:mass} shows the typical mass of objects, $M$, as a function
of redshift $z$; the labels $1\sigma$, $2\sigma$, and $3\sigma$ indicate
the amplitudes of initial density fluctuations in the universe from
which the objects form, on the assumption of the CDM fluctuations
spectrum \citep{bar01}. Fig.\ref{fig:temp} shows the downstream
temperature at the virial shock ($T_{\rm vir}$) and that at the LSS
shock ($T_s$) for the objects; $T_{\rm vir}$ is always larger than
$T_s$. The ratio $T_{\rm vir}/T_s$ indicates that ${\cal M}\ga 4$ for
the virial shock. For the LSS shock, ${\cal M}\gg 1$, because the gas
outside the shock is cold. Thus, equation~(\ref{eq:mag}) can be applied
to both shocks. In Fig.\ref{fig:B}, we present the strength of magnetic
fields ($B_c$) at a scale of $r_{\rm vir}$ for the collapsed objects. We
assume that the entire universe is reionised at $z=8$. Thus, for $z>8$,
magnetic fields are generated only at the virial shocks if
$T_s>10^4\rm\: K$. We assume that $B_c=B_f$ and plot the lines only when
$T_s>10^4\rm\: K$. The recombination is effective at $z>10$ for the
$3\sigma$ model. On the other hand, for $z<8$, the magnetic fields are
generated at the LSS shocks. We consider the compression of the fields
while the size of the gas sphere decreases from $r=r_{\rm ta}$ to
$r=r_{\rm vir}$, and thus we assume that $B_c=8^{2/3}\: B_f$. Moreover,
we plot the lines only for $T_{\rm vir}>2\times 10^5$~K, because below
this temperature, gas infall is suppressed by photoionisation heating
\citep{efs92,fur04}.

In Fig.\ref{fig:B}, the strength of magnetic fields generated by protons
reaches $\sim 10^{-8}$--$10^{-7}$~G and is very close to the values
observed in nearby galaxies and clusters of galaxies \citep[$\sim
10^{-6}$~G;][]{cla01,wid02,val04}. On a galactic scale, the gas sphere
may further contract to $r\ll r_{\rm vir}$ because of radiative
cooling. If magnetic fields are frozen in the gas, the strength exceeds
$\sim 10^{-6}$~G. However, if this happens, the magnetic energy exceeds
the thermal or kinetic energy of the gas. As a result, magnetic
reconnection may reduce the strength. In fact, the equipartition between
the magnetic energy density and the thermal or kinetic energy density
appears to be held in the Galaxy \citep{bec96}.

The strong magnetic fields shown in Fig.~\ref{fig:B} indicate that
strong amplification of magnetic fields, such as dynamo amplification,
is not required after formation of the galaxies and clusters. This is
consistent with the observations of galactic magnetic fields at $z\ga 2$
\citep{ath98}. Future observations of higher-redshift galaxies would
discriminate between our model and strong dynamo amplification models;
the latter predict much weaker magnetic fields at higher
redshifts. Moreover, since the predicted galactic magnetic fields are
comparable to those at present, they might have affected the formation
of stars in protogalaxies.  Fig.\ref{fig:B} also shows that our model
naturally explains the observational fact that the magnetic field
strengths of galaxies and galaxy clusters fall in a small range (a
factor of 10). Since our model predicts that magnetic fields are
generated around objects, magnetic fields in intergalactic space are not
required as the seed or origin of galactic magnetic fields.

Some recent numerical simulations indicated that gas is never heated to
$\sim T_{\rm vir}$ for less massive objects because radiative cooling is
efficient and the shocks forming at $\ga r_{\rm vir}$ are unstable
\citep{bir03,ker05}. This effect may be important for $M\la 10^{12}\:
M_\odot$. If it is correct, the generation of magnetic fields is only
effective for $z\la 5$ (the $3\sigma$ curve in Fig.~\ref{fig:mass}).
However, even for $M\la 10^{12}\: M_\odot$, multiple shocks may form at
the inner halo when the cold gas reaches there and collides each other
\citep{ker05}. Therefore, magnetic fields may be created there. The
details of these shocks could be studied by high-resolution numerical
simulations.

Although it would be difficult to directly observe the generation of
magnetic fields through the Weibel instability for distant high-redshift
galaxies, it would be easier for nearby clusters of galaxies. Since
clusters are now growing, LSS shocks should be developing outside of the
virial radii of the clusters \citep{min00,ryu03}. Since particles are
often accelerated at shocks, the synchrotron emission from the
accelerated particles could be observed with radio telescopes with high
sensitivity at low frequencies \citep*{kes04}. The total non-thermal
luminosity (synchrotron luminosity plus inverse Compton scattering with
cosmic microwave background [CMB] photons) is estimated as
\begin{equation}
 L_{\rm nt}\approx 
\frac{\epsilon}{r_{\rm ta}/u} \frac{f M}{\mpr}\frac{1}{2}\Vsh^2\:,
\end{equation}
where $\epsilon$ is the acceleration efficiency and $f$ is the gas
fraction of a cluster. If we assume $\epsilon=0.03$ \citep{tan98,mur00}
and $f=0.15$ \citep*{moh99}, the maximum value of $L_{\rm nt}$ is $\sim
10^{43}\rm\: erg\: s^{-1}$. Since the energy density of magnetic fields
($u_{\rm B}$) is smaller than that of the CMB ($u_\mathrm{CMB}$), most
of the non-thermal luminosity ($L_{\rm nt}$) is attributed to the
inverse Compton scattering such as $u_\mathrm{CMB} L_{\rm nt}/(u_{\rm
B}+u_\mathrm{CMB})$, which may have been detected in the hard X-ray band
\citep[$\ga 20$~keV;][]{fus04}. Thus, the synchrotron radio luminosity
is $u_{\rm B} L_{\rm nt}/(u_{\rm B}+u_\mathrm{CMB})\la 10^{39}\rm\:
erg\: s^{-1}$. Some of the diffuse radio sources observed in the
peripheral cluster regions (`radio relics') may be this emission
\citep{gov04}. Since the Weibel instability generates magnetic fields on
the plane of the shock front, the synchrotron emission should be
polarised perpendicular to the shock front \citep{med99}, which is
actually observed for some radio relics \citep{gov04}. The synchrotron
emission will tell us the positions of the LSS shocks, if they exist.
If magnetic fields are generated there, they should be observed only
downstream of the shock. This may be confirmed through Faraday rotation
measurements of radio sources behind the cluster for both sides of the
shock, if the coherent length of the fields sufficiently increases (see
Section 4).

\section{On the long-term evolution of magnetic fields}
\label{sec:uns}

The model proposed here has large uncertainties, especially on the {\it
very} long-term (say Gyr) evolution of magnetic fields.  The typical
time-scale of the Weibel instability until the saturation is given by
the inverse of the proton plasma frequency, $\omega_{\rm pp} \equiv
(4\pi \np e^2 / \mpr)^{1/2}$.  However, it is evident that this
time-scale is much smaller than the cosmological time-scale we have
considered in this paper.  The typical scale-length of the saturated
fields is given by the proton inertial length $c/\omega_{\rm pp}$, and
it is also much smaller than a kpc scale. Thus, there is a `missing
link' between the generated fields and those observed in galaxies and
clusters at present.

If the observed magnetic fields have their origin in the Weibel
instability, there should be another mechanism that takes over the
instability. If the saturation and current evolution model of
\citet{kat05} is applied to proton currents, the magnetic field strength
should decrease after the saturation as
\begin{equation}
\label{eq:magafter}
	B = \Bsat\ (R/\tilde{R})^{-1},
\end{equation}
where $R$ is the radius of a filament, and the filament radius at
saturation, $\tilde{R}$, is defined for protons ($\tilde{R} \sim 2
c/\omega_{\rm pp}$). This is because the current strength is limited to
the proton Alfv\'{e}n current $I'_A$ while the radius of current
filaments increases through mergers of the filaments.  Since $\tilde{R}
\sim 10^{10} \mathrm{cm}$ for $\np \sim 10^{-5} \mathrm{cm}^{-3}$ (at
$r\sim r_{\rm vir}$ for a cluster at $z\sim 0$), the magnetic field
strength should be reduced by a factor of $\sim 10^{11}$, if relation
\refp{eq:magafter} holds even for an long-term evolution to a kpc ($\sim
10^{21}$ cm) scale field.

One possibility to overcome this difficulty is that the currents are
carried in another form at later times.  At early times of the
evolution, the state of the plasma is in a kinetic regime in which the
characteristic scale of magnetic fields is comparable to or smaller than
the Larmor radius of particles. Probably, at later times, the former
would increase via current mergers, and would become larger than the
latter.  After this situation is realised, the straight currents
generated by the Weibel instability, which are limited to the Alfv\'{e}n
current, might be superseded by a kind of drift currents, and the plasma
might behave as a MHD fluid. In this case, mergers of the cylindrical
structures, which were originally the current filaments, would occur
through the reconnection process as usually considered for MHD fluids,
and therefore the magnetic field strength would not decrease
considerably unless the diffusion of magnetic fields owing to collisions
between charged particles or collisions with neutrals becomes
effective.\footnote{We note that `fast' reconnection \citep{pet64} may
occur on time-scales much shorter than those of the particle
collisions.}  This would allow the magnetic fields to survive for a long
time.  For example, in a MHD fluid, for the scale of $L\sim 10^{10}\rm\:
cm$ and the temperature of a few keV, the dissipation time-scale of
magnetic fields through the collisions between charged particles is
$t_{\rm diss}\sim 10^9$~yr \citep{spi62}, which is the dynamical
time-scale of a cluster. If current mergers make $L$ larger, $t_{\rm
diss}$ also becomes larger. The numerical simulations performed by
\citet{sil03} showed that magnetic fields do not much decay after the
saturation. This might reflect the transition to a MHD fluid. Moreover,
a MHD inverse cascade mechanism \citep[e.g.][]{vis01} could provide
another process to make larger-scale magnetic fields observed in
galaxies and clusters ($\ga$~kpc). Since the ionisation rate, $y$, is
fairly large (Section~\ref{sec:shockform}), the ambipolar diffusion of
magnetic fields (the collisions with neutrals) can be ignored
\citep[e.g. eq.~13--57 in][]{spi78}.  At any rate, studies about the
long-term evolution of magnetic fields are strongly encouraged.

\section{Conclusions}

In this paper, we showed that the Weibel instability can generate strong
magnetic fields in shocks around galaxies and clusters. The strength is
comparable to those observed in galaxies and clusters at present. The
mechanism could have worked even at $z\sim 10$. The results are based on
the assumption that the magnetic fields generated by the Weibel
instability are conserved for a long time. The validity of this
assumption must be confirmed in future studies.

\section*{Acknowledgments}

We are grateful to an anonymous referee for several suggestions that
improved this paper. We thank K.~Omukai, N.~Okabe, T.~Kudoh, K.~Asano,
and S.~Inoue for discussions. Y. F. is supported in part by a
Grant-in-Aid from the Ministry of Education, Culture, Sports, Science,
and Technology of Japan (14740175).

\appendix

\section{Dispersion relation of the Weibel instability}
\label{sec:AppendixA}
Here, we derive the dispersion relation of the Weibel instability in a
plasma which consists of some species (or populations) of charged
particles.  In the following, each species is denoted by a label `s' (s
= e for electron, p for proton, and so on), and the mass, charge, and
number density of the species are denoted by $\ms$, $\qs$ and $\ns$,
respectively.  We assume here that each species has a bi-Maxwellian
distribution
\begin{equation}
	f_0^{(\mathrm{s})}(\mathbf{v}) =
		\frac{\ns}{(2\pi)^{3/2} \alphas \sigmas^3}
		\exp \left[
			- \frac{v_x^2 + v_y^2}{2 \sigmas^2}
			- \frac{v_z^2}{2 \alphas^2 \sigmas^2}
		\right],
\end{equation}
where the thermal velocity is given by $\sigmas$ for $x$ or $y$
directions, and it is given by $\alphas \sigmas$ for $z$ direction.

If the thermal velocity in $z$ direction is larger than the other
directions \footnote{
Note that in this condition the direction of higher temperature
is opposite to that of other authors \cite[e.g.,][]{wei59, dav72}.
 Nevertheless, this would be
more reasonable for anisotropy at shock waves.
}, that is, $\alphas \ge 1$, the linear dispersion relation of the
Weibel mode, which is relevant to the $z$ component of the current
density, is given as
\begin{equation}
	\omega^2 - (ck)^2
		+ \sum_\mathrm{s} \omegaps^2 \left[
			\alphas^2 \zetas Z(\zetas)
			+ \alphas^2 - 1
		\right]
	= 0,
	\label{eq:Weibel_dispersion}
\end{equation}
where
\begin{equation}
	\omegaps \equiv \sqrt{\frac{4\pi\ns\qs^2}{\ms}},
	\qquad
	\zetas \equiv \frac{\omega}{k \sqrt{2} \sigmas},
\end{equation}
and $Z(\zeta)$ is the plasma dispersion function defined as
\begin{equation}
	Z(\zeta)
		\equiv \frac{1}{\sqrt{\pi}} \int_{-\infty}^\infty
			\frac{1}{z - \zeta} e^{-z^2} dz.
\end{equation}

It is shown that the dispersion relation \refp{eq:Weibel_dispersion} can
have purely positive-imaginary solution of $\omega$, i.e., purely
growing mode.  Since the growth rate becomes zero at the maximum wave
number of the unstable mode, $\kmax$, we can set $\omega = 0$ and
$\zetas Z(\zetas) = 0$ at $k=\kmax$ in Eq.\refp{eq:Weibel_dispersion} to
obtain
\begin{equation}
	k_\mathrm{max}^2
	=
	\frac{1}{c^2} \sum_\mathrm{s} \omegaps^2 (\alphas^2 -1).
	\label{eq:kmax}
\end{equation}
It is evident that there is no unstable mode if $k_\mathrm{max}^2 \le
0$, while unstable modes exist if $k_\mathrm{max}^2 > 0$.  Explicitly,
the condition for instability is given by
\begin{equation}
	\sum_\mathrm{s} \omegaps^2 (\alphas^2 -1) > 0.
	\label{eq:instability_condition}
\end{equation}
It should be noted that, in an electron-proton plasma for example, even
when the electron distribution is completely isotropic
($\alpha_\mathrm{e}=1$) at the initial time, unstable modes exist if the
proton distribution is anisotropic ($\alpha_\mathrm{p} > 1$).

The above results show that even if the Mach number of a shock is
relatively small ($\alpha_\mathrm{e}\approx 1$), the magnetic fields
generated by the anisotropy of proton distribution slowly grow and reach
the saturation value given by equation~(\ref{eq:mag}) \citep{kat05}.

\label{lastpage}


\begin{thebibliography}{99}

\bibitem[\protect\citeauthoryear{Alfv\'{e}n}{1939}]{alf39} 
Alfv\'{e}n H., 1939, Phys. Rev., 55, 425 

\bibitem[\protect\citeauthoryear{Athreya et 
al.}{1998}]{ath98} Athreya R.~M., Kapahi V.~K., McCarthy 
P.~J., van Breugel W., 1998, A\&A, 329, 809 

\bibitem[\protect\citeauthoryear{Barkana \& Loeb}{2001}]{bar01} Barkana
						   R., Loeb A., 2001,
						 Physics Reports, 349, 125

\bibitem[\protect\citeauthoryear{Beck et al.}{1996}]{bec96} 
Beck R., Brandenburg A., Moss D., Shukurov A., Sokoloff D., 1996, ARA\&A, 
34, 155 
 
\bibitem[\protect\citeauthoryear{Bertschinger}{1985}]{ber85} 
Bertschinger E., 1985, ApJS, 58, 39 

\bibitem[\protect\citeauthoryear{Birnboim \&
Dekel}{2003}]{bir03} Birnboim Y., Dekel A., 2003, MNRAS, 345, 
349 

\bibitem[\protect\citeauthoryear{Bryan \& 
Norman}{1998}]{bry98} Bryan G.~L., Norman M.~L., 1998, ApJ, 
495, 80 

\bibitem[\protect\citeauthoryear{Cen \& 
Ostriker}{1999}]{cen99} Cen R., Ostriker J.~P., 1999, ApJ, 
519, L109 

\bibitem[\protect\citeauthoryear{Clarke, Kronberg, \&
	   B{\"o}hringer}{Clarke et al.}{2001}]{cla01} Clarke, T.~E.,
	   Kronberg,
P.~P., \& B{\" o}hringer, H.\ 2001, ApJ, 547, L111 

\bibitem[\protect\citeauthoryear{Dav{\' e} et 
al.}{2001}]{dav01} Dav{\' e} R., et al., 2001, ApJ, 552,
	   473 
\bibitem[\protect\citeauthoryear{Davidson et al.}{1972}]{dav72}
	   Davidson, R.~C., Hammer, D.~A.,
			      Haber, I., Wagner, C.~E.\ 1972,
			      Phys. Fluids, 317, 333

\bibitem[\protect\citeauthoryear{Efstathiou}{1992}]{efs92} Efstathiou,
	   G.\ 1992, MNRAS, 256, 43P

\bibitem[\protect\citeauthoryear{Frederiksen et al.}{2004}]{fre04}
	   Frederiksen, J.~T.,
Hededal, C.~B., Haugb{\o}lle, T., Nordlund, {\AA}.\ 2004, ApJ, 608, 
L13 
 
\bibitem[\protect\citeauthoryear{Fried}{1959}]{fri59} Fried, B.~D.\
	   1959, Phys. Fluids, 2, 83

\bibitem[\protect\citeauthoryear{Furlanetto \& Loeb}{2004}]{fur04}
	   Furlanetto,
S.~R., Loeb, A.\ 2004, ApJ, 611, 642

\bibitem[\protect\citeauthoryear{Fusco-Femiano et al.}{2004}]{fus04}
	   Fusco-Femiano,
R., Orlandini, M., Brunetti, G., Feretti, L., Giovannini, G., Grandi, P., 
 Setti, G.\ 2004, ApJ, 602, L73 

\bibitem[\protect\citeauthoryear{Govoni \& Feretti}{2004}]{gov04}
	   Govoni, F., Feretti, L.\ 2004,
			      Int. J. Mod. Phys. D13 1549

\bibitem[\protect\citeauthoryear{Gunn \& Gott}{1972}]{gun72} Gunn,
	   J.~E., Gott,
J.~R.~I.\ 1972, ApJ, 176, 1 

\bibitem[\protect\citeauthoryear{Kato}{2005}]{kat05} Kato, T.~N.,\ 2005,
	   Phys. Plasmas, 12, 080705

\bibitem[\protect\citeauthoryear{Kazimura et al.}{1998}]{kaz98}
	   Kazimura, Y., Sakai,
J.~I., Neubert, T., Bulanov, S.~V.\ 1998, ApJ, 498, L183 

\bibitem[\protect\citeauthoryear{Kere\v{s} et al.}{2005}]{ker05}
Kere\v{s}, D., Katz, N., Weinberg, D.~H., Dav\'{e}, R.\ 2005,
	   MNRAS in press (astro-ph/0407095)

\bibitem[\protect\citeauthoryear{Keshet, Waxman, \&
	   Loeb}{Keshet et al.}{2004}]{kes04} Keshet, U.,
			      Waxman, E.,
 Loeb, A.\ 2004, ApJ, 617, 281 

\bibitem[\protect\citeauthoryear{Medvedev et al.}{2005}]{med05}
	   Medvedev, M.~V.,
Fiore, M., Fonseca, R.~A., Silva, L.~O., Mori, W.~B.\ 2005, 
ApJ, 618, 
L75 

\bibitem[\protect\citeauthoryear{Medvedev \& Loeb}{1999}]{med99}
	   Medvedev, M.~V., 
Loeb, A.\ 1999, ApJ, 526, 697 

\bibitem[\protect\citeauthoryear{Miniati et al.}{2000}]{min00} Miniati,
	   F., Ryu, D.,
Kang, H., Jones, T.~W., Cen, R.,  Ostriker, J.~P.\ 2000, ApJ, 542,
			      608 

\bibitem[\protect\citeauthoryear{Mohr, Mathiesen, \& Evrard}{Mohr et
	   al.}{1999}]{moh99} Mohr, J.~J., Mathiesen,
B., Evrard, A.~E.\ 1999, ApJ, 517, 627 

\bibitem[\protect\citeauthoryear{Muraishi et al.}{2000}]{mur00}
	   Muraishi, H., et al.\
2000, A\&A, 354, L57 

\bibitem[\protect\citeauthoryear{Nishikawa et al.}{2003}]{nis03}
	   Nishikawa, K.-I.,
Hardee, P., Richardson, G., Preece, R., Sol, H., Fishman, G.~J.\ 2003, 
ApJ, 595, 555 

\bibitem[\protect\citeauthoryear{Okabe \& Hattori}{2003}]{oka03} Okabe,
	   N., \&
Hattori, M.\ 2003, ApJ, 599, 964 

\bibitem[\protect\citeauthoryear{Peebles}{1980}]{pee80} Peebles,
	   P.~J.~E.\ 1980, Large-Scale
			      Structure of the
	Universe (Princeton Univ. Press; Princeton)

\bibitem[\protect\citeauthoryear{Petschek}{1964}]{pet64}
Petschek, H.~E. 1964, in The Physics of Solar Flares, ed. W. N. Hess
						   (Washington, DC:
						   NASA), 425

\bibitem[\protect\citeauthoryear{Ryu et al.}{2003}]{ryu03} Ryu, D.,
	   Kang, H., Hallman,
E., Jones, T.~W.\ 2003, ApJ, 593, 599 

\bibitem[\protect\citeauthoryear{Schlickeiser \& Shukla}{2003}]{sch03}
	   Schlickeiser,
R., Shukla, P.~K.\ 2003, ApJ, 599, L57 

\bibitem[\protect\citeauthoryear{Shapiro \& Kang}{1987}]{sha87} Shapiro,
	   P.~R., \&
Kang, H.\ 1987, ApJ, 318, 32 

\bibitem[\protect\citeauthoryear{Silva et al.}{2003}]{sil03} Silva,
	   L.~O., Fonseca,
R.~A., Tonge, J.~W., Dawson, J.~M., Mori, W.~B., Medvedev, M.~V.\ 
2003, 
ApJ, 596, L121 

\bibitem[\protect\citeauthoryear{Spitzer}{1962}]{spi62} Spitzer, Jr.
	   L. 1962, Physics of Fully Ionized Gases (New York: Wiley)

\bibitem[\protect\citeauthoryear{Spitzer}{1978}]{spi78} Spitzer, Jr.
	   L. 1978, Physical Processes in
			      the Interstellar Medium (New York: Wiley)

\bibitem[\protect\citeauthoryear{Susa et al.}{1998}]{sus98} Susa, H.,
	   Uehara, H.,
Nishi, R., Yamada, M.\ 1998, Progress of Theoretical Physics, 100, 63 

\bibitem[\protect\citeauthoryear{Tanimori et al.}{1998}]{tan98}
	   Tanimori, T., et al.\
1998, ApJ, 497, L25 

\bibitem[\protect\citeauthoryear{Vall{\' e}e}{2004}]{val04} Vall{\' e}e,
	   J.~P.\ 2004,
New Astronomy Review, 48, 763 

\bibitem[Vishniac \& Cho(2001)]{vis01} Vishniac, E.~T., 
Cho, J.\ 2001, ApJ, 550, 752 

\bibitem[\protect\citeauthoryear{Wei et al.}{2002}]{wei02} Wei, M.~S.,
	   et al.\ 2002,
	Central Laser Facility (UK) Annual Report 
(http://www.clf.rl.ac.uk/Reports/2001-2002/pdf/10.pdf)

\bibitem[\protect\citeauthoryear{Weibel}{1959}]{wei59} Weibel, E.~S.\
	   1959, Phys. Rev. Lett., 2,
			      83

\bibitem[\protect\citeauthoryear{Widrow}{2002}]{wid02} Widrow, L.~M.\
	   2002, Reviews of
Modern Physics, 74, 775 

\end{thebibliography}
\end{document}